\documentstyle[apjfonts,emulateapj,epsf,psfig]{article}

\newcommand\FX{ergs~s$^{-1}$~cm$^{-2}$}
\newcommand\LX{ergs~s$^{-1}$}
\newcommand\ftool{{\tt FTOOL}}

\newcommand{\degrees}{\mbox{$^{\circ}$}}

\newcommand\asca{{\it ASCA}}

\newcommand\rosat{{\it ROSAT}}

\newcommand\bepposax{{\it Beppo-SAX}}

\newcommand\einstein{{\it Einstein}}

\def\ltsima{$\; \buildrel < \over \sim \;$}
\def\simlt{\lower.5ex\hbox{\ltsima}}
\def\gtsima{$\; \buildrel > \over \sim \;$}
\def\simgt{\lower.5ex\hbox{\gtsima}}
\newcommand\psrG{PSR~J1119$-$6127}
\newcommand\snrG{G292.2$-$0.5}
\newcommand\axG{AX~J1119.1$-$6128.5}
\newcommand\irasG{IRAS~J11169$-$6111}
\newcommand\psra{PSR~B1046$-$58}

\slugcomment{Accepted for publication by The Astrophysical Journal February 2, 2001}

\lefthead{Pivovaroff, Kaspi, Camilo, Gaensler \& Crawford}
\righthead{PSR J1119$-$6127 \& SNR G292.2$-$0.5}

\begin{document}
\title{X-ray Observations of the New Pulsar--Supernova Remnant System PSR~J1119$-$6127 and
SNR~G292.2$-$0.5}

\author{M. J. Pivovaroff,\altaffilmark{1,6}
V. M. Kaspi,\altaffilmark{2,1,3}
F. Camilo,\altaffilmark{4}
B. M. Gaensler,\altaffilmark{1,5} and
F. Crawford\altaffilmark{1,7}}

\altaffiltext{1}{Department of Physics and Center for Space Research,
Massachusetts Institute of Technology, Cambridge, MA~02139}
\altaffiltext{2}{Department of Physics, Rutherford Physics Building,
McGill University, 3600 University Street, Montreal, Quebec, H3A~2T8,
Canada}
\altaffiltext{3}{Alfred P. Sloan Research Fellow}
\altaffiltext{4}{Columbia Astrophysics Laboratory, Columbia
University, 550 W. 120th Street, New York, NY~10027}
\altaffiltext{5}{Hubble Fellow}
\altaffiltext{6}{Current address: Therma-Wave Inc., 1250 Reliance Way,
Fremont, CA 94539; mpivovar@thermawave.com}
\altaffiltext{7}{Current address: Management and Data Systems Division,
Lockheed Martin Corporation, PO Box 8048, Philadelphia, PA 19101}

\authoremail{mjp@space.mit.edu, vkaspi@physics.mcgill.ca,
fernando@astro.columbia.edu, bmg@space.mit.edu, crawford@space.mit.edu}

\begin{abstract}
\psrG\ is a recently discovered 1700-year-old radio pulsar that
has a very high inferred surface dipolar magnetic field. We
present a detailed analysis of a pointed \asca\ observation and
archival \rosat\ data of \psrG\ and its surroundings. Both data
sets reveal extended emission coincident with the newly-discovered
radio supernova remnant G292.2--0.5, reported in a companion paper
by Crawford et al. A hard point source, offset $\sim$1$\farcm$5
from the position of the radio pulsar, is seen with the \asca\
GIS. No pulsations are detected at the radio period with a pulsed
fraction upper limit of 61\% (95\% confidence). The limited
statistics prevent a detailed spectral analysis, although a
power-law model with photon index $\Gamma \approx 1-2$ describes
the data well. Both the spectral model and derived X-ray
luminosity are consistent with those measured for other young
radio pulsars, although the spatial offset renders an
identification of the source as the X-ray counterpart of the
pulsar uncertain.

\end{abstract}

\section{Introduction}

X-ray observations of young rotation-powered pulsars offer a
unique opportunity to resolve several fundamental questions about
neutron stars and supernova remnants. Studying the spectrum and
morphology of the pulsar's synchrotron nebula (or plerion; Weiler
\& Panagia 1978) \nocite{wp78} is crucial for determining basic
properties about the relativistic pulsar wind and probing the
density of the surrounding medium. Measuring the temperature of
the cooling surface of the pulsar provides a way to understand the
thermal evolution of neutron stars and to constrain the equation
of state. Additionally, X-ray observations of pulsars are also
important for searching for the remnant of the supernova that
created the pulsar. Direct studies of the remnant are useful for
numerous reasons, including measurement of the elemental
abundances and determination of the shock conditions in the
supernova remnant (SNR) through spectral analysis. Associations
between pulsars and SNRs can provide independent distance and age
estimates for both objects, and with a statistically significant
sample of associations, constraints on the birth properties of
neutron stars, including initial period, magnetic field, and
velocity distribution, can be found.

\psrG\ was discovered by Camilo et al. (2000) \nocite{ckl+00} in
an on-going survey for pulsars in the Galactic plane using the
Parkes 64-m radio telescope in Australia. Its spin period of
$P=0.4$~s is long compared to those of young Crab-like pulsars,
but its period derivative of $\dot P = 4 \times 10^{-12}$ is
extremely large. The characteristic age calculated from these
parameters is only $\tau_{c} \equiv P/2\dot{P} = 1600$~yr. In
addition to its extreme youth, \psrG\ also has a very large
inferred surface dipole magnetic field strength, $B \simeq 3.2
\times 10^{19} (P\dot{P})^{1/2}~{\rm G} = 4 \times 10^{13}$~G.

\psrG\ is also noteworthy as it is one of only five pulsars that
has an accurately measured second period derivative, $\ddot{P}$,
which allows determination of the braking index. In many ways,
\psrG\ is similar to PSR~B1509$-$58, another long-period young
pulsar for which the braking index is measured.
Table~\ref{tab:j1119_astrometry} presents the spin parameters and
derived quantities for both of these pulsars. In the same way
that PSR~B0540$-$69 and the Crab pulsar establish the existence of
a class of rapidly-spinning young pulsars with large spin-down
luminosities, \psrG\ and PSR~B1509$-$58 suggest the existence of a
class of equally young pulsars but with higher magnetic fields,
longer periods, and much lower spin-down luminosities.

{\scriptsize
\begin{deluxetable}{l r r}
\tablewidth{0pt} \tablecaption{Astrometric and spin parameters for
PSRs~J1119$-$6127 and B1509$-$58 \label{tab:j1119_astrometry} }
\tablehead{
 \colhead{Parameter} & \colhead{\psrG}  &
     \colhead{PSR B1509$-$58}
} \startdata Right ascension (J2000)&
   $11^{\rm h} \; 19^{\rm m} \; 14\fs30$ &
   $15^{\rm h} \; 13^{\rm m} \; 55\fs62$ \nl
Declination (J2000)&
   $-61^{\circ} \; 27' \; 48\farcs5$ &
   $-59^{\circ} \; 08' \; 09\farcs0$ \nl
Period, $P$ (ms)&
   407.64  &  150.66  \nl
Period derivative, $\dot{P}$  &
   $4.023 \times 10^{-12}$ & $1.537\times 10^{-12}$   \nl
Second period derivative, $\ddot{P}$ &
   $3.59 \times 10^{-23}$ & $1.31\times 10^{-23}$   \nl
Epoch of period (MJD)             & 51173.0    &  48355.0 \nl
Braking index &
   $2.91 \pm 0.05$ &  $2.837 \pm 0.001$  \nl
Dispersion measure, DM (pc cm$^{-3}$)&
   707   & 253  \nl
Characteristic age, $\tau_{c}$ (yr)              &
   1606 &  1554 \nl
Spin-down luminosity, $\dot{E}$ (ergs s$^{-1}$)  &
   $2.3 \times 10^{36}$  & $1.8 \times 10^{37}$ \nl
Magnetic dipole field strength, $B$ (G) &
   $4.1 \times 10^{13}$ & $1.5 \times 10^{13}$ \nl
Reference  & Camilo et al. (2000) & Kaspi et al. (1994)
\nocite{kms+94} \nl
\enddata
\end{deluxetable}
}

After the pulsar's discovery, relevant publicly accessible data
archives were searched for observations with \psrG\ in their field
of view. A survey of the Galaxy performed with the Molonglo
Observatory Synthesis Telescope (MOST) at 843~MHz (Green et al.
1999) \nocite{gcly99} revealed a faint ring-like shell,
$\sim$15\arcmin\ in extent and approximately centered on the
position of the pulsar. A brief \rosat\ PSPC observation shows
X-ray emission coincident with a portion of the radio shell.
Recently, radio interferometric observations of \psrG\ and its
vicinity have been made with the Australia Telescope Compact Array
(ATCA) (Crawford 2000; Crawford et al. 2001).
\nocite{cra00,cgk+00} The radio data confirm the existence of the
shell-like emission and show that it has a non-thermal spectrum,
characteristic of a shell SNR. In this paper we present a detailed
analysis of both a pointed \asca\ and the serendipitous \rosat\
observation.

\section{Observations}

The \asca\ X-ray telescope (Tanaka, Inoue, \& Holt 1994)
\nocite{tih94} was used to observe \psrG\ during a 36 hour period
spanning 1999 August 14--15 as part of the AO-7 Guest Observer
program (sequence number 57040000). To achieve the time resolution
necessary for pulsation searches, the GIS (Gas Imaging
Spectrometer) was operated in a non-standard mode. A time
resolution of 0.488~ms or 3.91~ms (depending on the telemetry
rate) was achieved by sacrificing information about the time
characteristics (or ``Risetime'') of detected events, one way to
differentiate between background and celestial X-ray photons (see
$\S$\ref{sec:j1119_image_asca} and \ref{sec:j1119_timing} for a
discussion of other ramifications of this operation mode).

As a compromise between having a field-of-view (FOV) wide
enough to encompass a large fraction of the shell and obtaining
potentially useful spectroscopic information from the CCDs, the
SIS (Solid-state Imaging Spectrometer) was operated in two-chip
mode, providing an imaging area $22\arcmin \times 11\arcmin$. The
data were analyzed using the standard (i.e. {\tt REV 2}) screening
criteria suggested in {\it The ASCA Data Reduction
Guide.\footnote{
http://legacy.gsfc.nasa.gov/docs/asca/abc/abc.html.}} The
resulting effective exposure times are 37~ks (GIS) and 34~ks
(SIS).

The \rosat\ X-ray telescope (Tr\"{u}mper 1983) \nocite{tru83}
serendipitously observed the field around \psrG\ with the Position
Sensitive Proportional Counter (PSPC) on 1996 August 14 during an
11-ks pointed observation of NGC~3603 (sequence number
RP900526N00). After retrieving the data from the NASA-maintained
HEASARC archive, we used the already-processed and filtered data
for our subsequent analysis. In the observation, \psrG\ is
located 31\arcmin\ away from the optical axis, and due to
vignetting and obscuration from the mirror support structure, the
effective exposure time for a 15\arcmin\ diameter circle centered
on the pulsar is between 6.4--8.9~ks.

\section{Data Reduction}
\subsection{Image Analysis}
\subsubsection{{\it ASCA}}
\label{sec:j1119_image_asca}

Flat-fielded images were generated by aligning and co-adding
exposure-corrected images from pairs of instruments. As discussed
by Pivovaroff, Kaspi \& Gotthelf (2000), \nocite{pkg00} great
care must be taken (particularly for the GIS) to remove detector
artifacts (e.g., the window support grid) and the
diffuse X-ray background (XRB). The software employed includes
the recently released \ftool\ {\tt mkgisbgd}.
\footnote{http://heasarc.gsfc.nasa.gov/docs/asca/mkgisbgd/mkgisbgd.html.}
For a complete discussion of the steps taken to make this software
compatible with this observation, see Chapter 7 of Pivovaroff
(2000).

\begin{figure*}
\centerline {\epsfbox{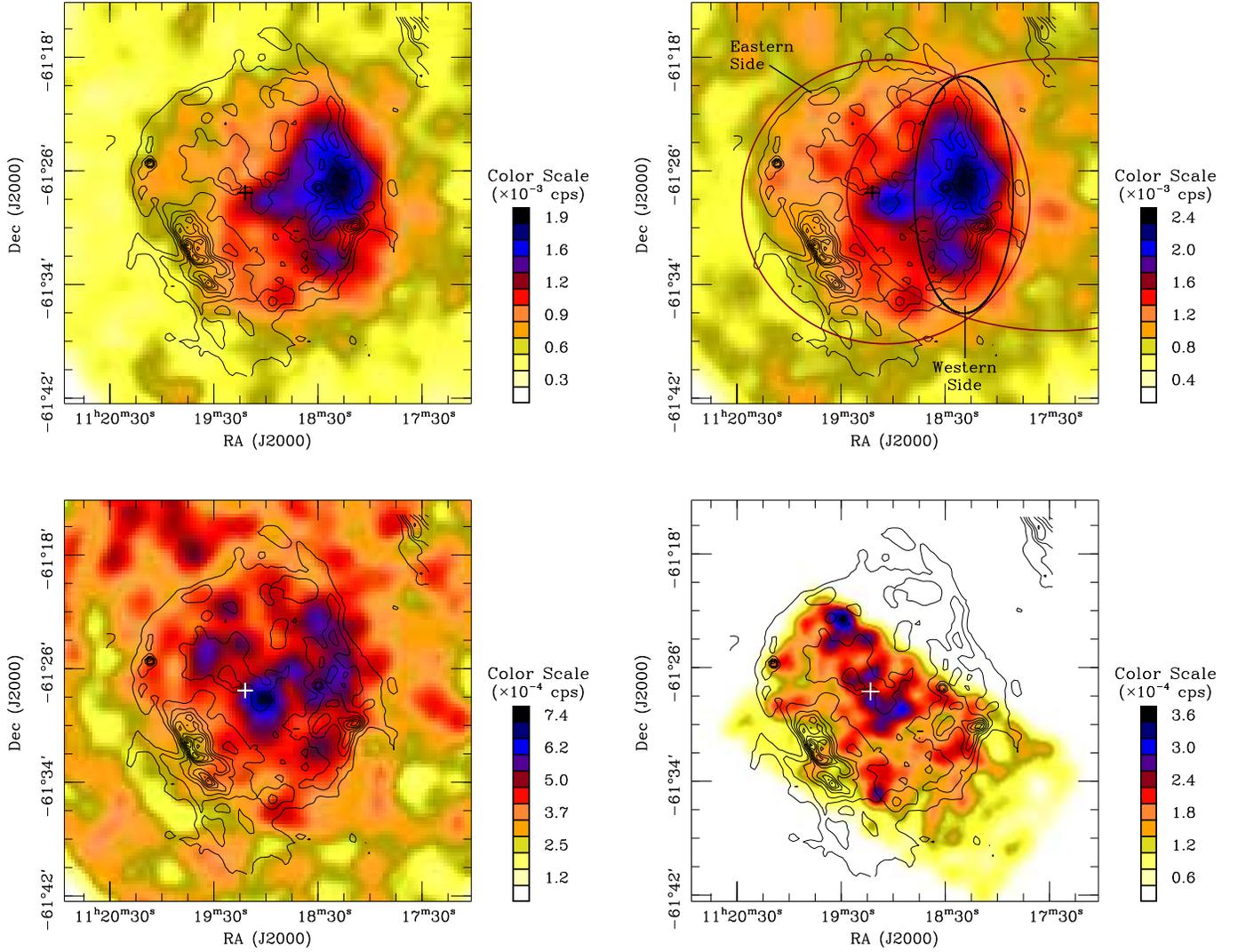}}
\caption{\asca\ images of the field around \psrG.  Images are made
in three separate energy bands: soft-band (0.8--3.0~keV),
hard-band (3.0--10.0~keV), and broad-band (0.8--10.0~keV). GIS
images are shown for each band: soft ({\it top left}), hard ({\it
bottom left}), and broad ({\it top right}).  For comparison, the
SIS hard-band image is also shown ({\it bottom right}). In each
image, the location of \psrG\ is marked by a cross. The contours
are from 1.4 GHz ATCA observations and range from 5\% to 95\% of
the maximum value (39~mJy~beam$^{-1}$) in increments of 15\%
(Crawford et al. 2001). The color bars in each plot indicate
counts~s$^{-1}$~arcmin$^{-2}$. The X-ray emission, classified here
as the previously unknown SNR \snrG, roughly traces the radio
morphology in the broad- and soft-bands, although significant
enhancement is seen on the western (right) side of the SNR below
3~keV.  A hard point-like source is clearly evident in the GIS
image.  The GIS broad-band image also shows the two regions used
to extract spectra of the SNR.}
\label{fig:j1119_asca}
\end{figure*}

After rebinning the SIS data by a factor of four
(6$\farcs$73~pixel$^{-1}$), data from both instruments were
smoothed with a kernel representing the point spread function
(PSF) of the X-ray telescope (XRT) and detector combination. We
approximate this function with a Gaussian of $\sigma = 30\arcsec$
for the SIS and $\sigma = 45\arcsec$ for the GIS. Finally, we
correct the astrometric position of the smoothed images for known
errors in the pointing solution using the \ftool\
{\tt offsetcoord}\footnote{Refer
 to http://heasarc.gsfc.nasa.gov/docs/asca/coord/updatecoord.html for
details.}.

X-ray events were filtered by energy to make images in three
different bands: soft (0.8--3.0~keV), hard (3.0--10.0~keV), and
broad (0.8--10.0~keV). Figure~\ref{fig:j1119_asca} shows GIS
images for all three band (soft [{\it top left}], hard [{\it
bottom left}], and broad [{\it top right}]). Given the
similarities between the two instruments, only the hard-band SIS
image is shown ({\it bottom right}). In each plot, the radio
position of \psrG\ is marked by a cross. Each plot also displays
the contours from the recent 1.4-GHz ATCA radio observations of
the field around the pulsar (Crawford et al. 2001). Both
instruments show significant emission from a nearly circular
region that roughly matches the radio SNR morphology. On the
basis of this and other evidence presented below, we classify this
extended emission as the previously unidentified X-ray-bright
supernova remnant \snrG. The right (western) side of the SNR
exhibits marked enhancements in X-ray flux at energies below
3~keV, while at higher energies the emission is relatively uniform
throughout the SNR.

In addition to the X-ray bright SNR, the GIS shows evidence for a
hard point-like source in the middle of \snrG. Following the
procedure of Pivovaroff, Kaspi, and Camilo (2000), \nocite{pkc00}
we estimate the significance of this detection by comparing the
number of photons detected in a small aperture centered on the
source with those from a concentric annulus used to estimate the
local background. We calculate a significance of $\sigma \gtrsim
4.0$ for the 158 background-subtracted photons detected with
energies between 3.0--10.0~keV. To explore the possibility that
the source is an instrumental artifact, we examined individual
images from GIS--2 and GIS--3. Emission is present in both
detectors at the above position, verifying the celestial nature of
the source.

\begin{figure*}
\centerline {\epsfbox{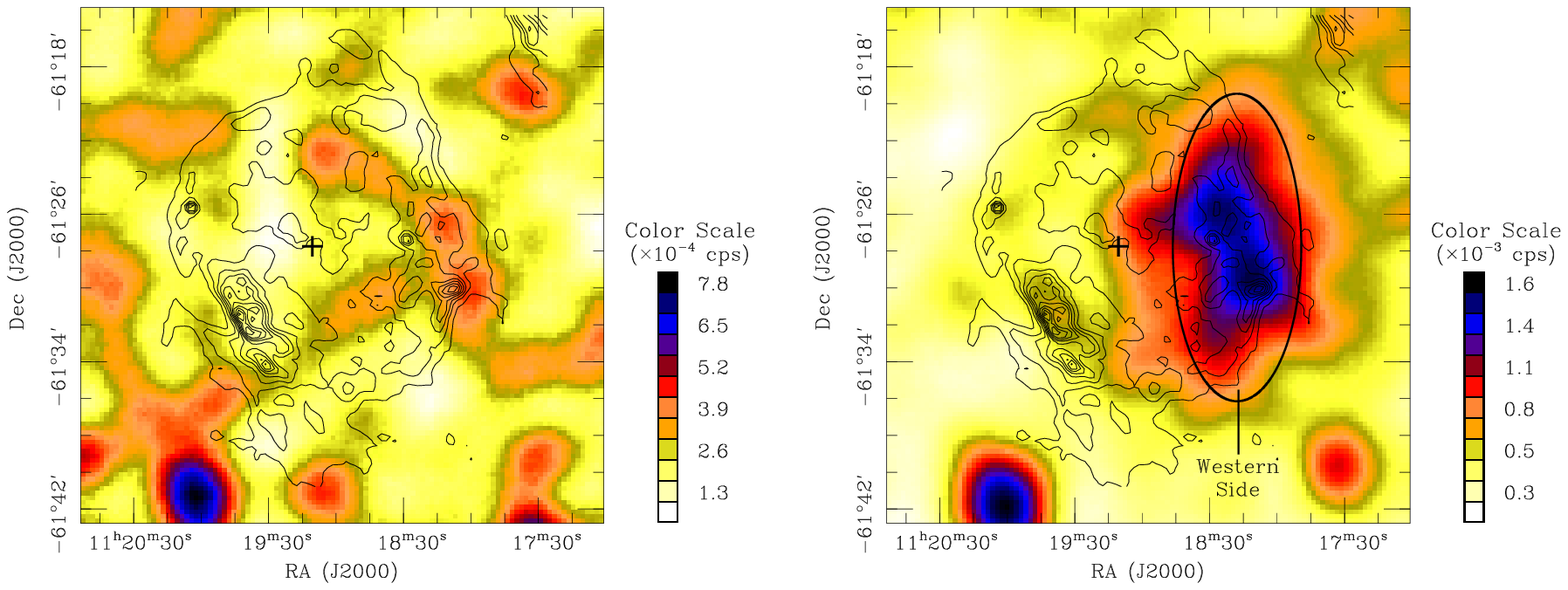}}
\caption{\rosat\ PSPC images of the field around \psrG. In each
image, the location of \psrG\ is marked by a cross. The contours
are from 1.4 GHZ  ATCA radio observations, as in
Figure~\ref{fig:j1119_asca}. The color bars in each plot indicate
counts~s$^{-1}$~arcmin$^{-2}$.  {\it Left:} The soft-band
(0.1--0.4~keV) image shows no emission coincident with either the
radio contours or the pulsar.  {\it Right:} The hard-band
(0.5--2.0~keV) image clearly shows emission from the western side
of the SNR and has the same morphology and intensity distribution
as the soft-band GIS image (Fig.~\ref{fig:j1119_asca} [{\it top
left}]). The heavy ellipse marks the region used to extract a
spectrum from the SNR.  No X-ray counterpart to \psrG\ is visible
in either band.}
\label{fig:j1119_rosat}
\end{figure*}

No point-like source is seen in the hard-band SIS image, although
there is enhanced emission at the source position derived from the
GIS data. Using aperture and annulus radii ($2\farcm0$,
$3\farcm25$, and $4\farcm5$, respectively) appropriate for the
smaller SIS$+$XRT PSF, we calculate a significance of $\sigma
\gtrsim 3.5$ for the 59 background-subtracted photons detected
between 3.0--10.0~keV. The lack of a more significant point-like
feature in the SIS when one is present in the GIS is not uncommon
and probably results from the larger effective area of the GIS at
high energies (Gotthelf, priv. comm.).

Fitting a two-dimensional Gaussian to the GIS source distribution,
we measure a position (J2000) of ${\rm R.A.} = 11^{\rm h}~19^{\rm
m}~03\fs4$, ${\rm Decl.} = -61\degrees~28^{\prime}$~30\arcsec\
with an uncertainty of $\sim$10\arcsec. We classify this source as
\axG. The absolute pointing uncertainty of the GIS, after making
the known corrections mentioned above, has an error radius of
24\arcsec\ at the 90\% confidence level (Gotthelf et al. 2000a).
\nocite{got00} The source is located 87\arcsec\ away from \psrG,
more than twice the combined positional errors. However, we note
that in at least one documented case, the \asca-measured position
of a source was 80\arcsec\ from the well established position
(Gotthelf et al. 2000a). Thus, while the offset of the GIS source
from the pulsar is larger than expected, it does not preclude the
possibility that the source is the X-ray counterpart of \psrG,
especially if other evidence supports the association.

\subsubsection{{\it ROSAT}}
Using standard software available from HEASARC, we produced
flat-fielded images with 15\arcsec\ pixels in the standard \rosat\
soft ($0.1-0.4$ keV) and hard ($0.5-2.0$ keV) bands. The
flat-fielded images were then smoothed with a Gaussian kernel
($\sigma = 1\farcm25$) to approximate the PSF of the telescope at
the off-axis position of the pulsar. Figure~\ref{fig:j1119_rosat}
shows the resultant images for the soft ({\it left\/}) and hard
({\it right\/}) bands. A cross marks the position of \psrG\ and
the contours are from the ATCA observations (Crawford et al.
2001). \nocite{cra00,cgk+00} No emission from the SNR is
detected in the soft band. However, in the hard band, emission
coincident with the western side of the radio shell is strongly
detected. No X-rays are detected from the pulsar in either band.
The morphology of the hard \rosat\ band (0.5--2.0~keV;
Fig.~\ref{fig:j1119_rosat} [{\it right\/}]) agrees well with that
of the soft GIS band (0.8--3.0~keV; Fig.~\ref{fig:j1119_asca}
[{\it top left}]).

The agreement between the two telescopes in principle offers the
chance to check the absolute pointing of \asca. Due to its large
FOV (2\degrees\ diameter) and its soft-energy sensitivity, the
PSPC usually detects emission from several nearby stars in any
given pointing. The positions of well resolved point sources were
checked for optical counterparts using the SIMBAD
catalog.\footnote{ http://cdsweb.u-strasbg.fr/Simbad.html.} Four
bright stars were found, and the mean offset between the optical
and X-ray positions was 18\arcsec. Thus, we take the absolute
position uncertainty of these PSPC data to be 18\arcsec. Next, we
cross-correlated the morphology between the PSPC and GIS data.
Unfortunately, the $\sim$arcminute spatial resolution of both
images, combined with the different responses of both instruments,
makes a detailed alignment check impossible. While an offset of
more than $\sim$1\arcmin\ between the \asca\ and \rosat\
observations can be ruled out, this does not represent an
improvement on the \asca\ positional uncertainty of $0\farcm5$
already discussed.

\subsection{Timing Analysis}
\label{sec:j1119_timing}

Pulsations from \psrG\ were searched for by extracting GIS events
from a circular region centered on the position of \axG. After the
arrival times were reduced to the barycenter using the \ftool\
{\tt timeconv}, the data were folded using 10 phase bins using the
radio ephemeris corresponding to the mean MJD of the \asca\
observation. Although the \asca\ observation of \psrG\ occurred
very close to the time the pulsar glitched (Camilo et al. 2000),
given the small differences in the pre- and post-glitch values of
$P$ and $\dot{P}$ and the total time span of our observation, the
glitch is irrelevant to our analysis.

{\tiny
\begin{deluxetable}{lcc ccc c cc c}
\tablecolumns{10} \tablewidth{6in} \tablecaption{Spectral fit
parameters for SNR G292.2$-$0.5 \label{tab:j1119_spectral}}
\tablehead{ \colhead{} & \colhead{} & \colhead{} &
\multicolumn{3}{c}{Western Side} & \colhead{} &
\multicolumn{2}{c}{Eastern Side} & \colhead{} \\ \cline{4-6}
\cline {8-9} \colhead{} & \colhead{$kT$} &  \colhead{Photon} &
\colhead{$N_{H}$} & \colhead{Norm\tablenotemark{a}} &
\colhead{Norm\tablenotemark{b}} &\colhead{} &\colhead{$N_{H}$} &
\colhead{Norm\tablenotemark{a}} & \colhead{}  \\
\colhead{Model} & \colhead{(keV)} & \colhead{Index} &
\colhead{($10^{22}$ cm$^{-2}$)} & \colhead{($10^{-3}$)} &
\colhead{($10^{-3}$)} & \colhead{} & \colhead{($10^{22}$
cm$^{-2}$)} & \colhead{($10^{-3}$)} & \colhead{$\chi^{2}$/dof} }
\startdata Power law &
  \nodata & $2.3 \pm 0.1$ &
  $0.28^{+0.07}_{-0.06}$ & $0.97^{+0.13}_{-0.11}$ & $0.62^{+0.12}_{-0.11}$ &&
  $1.6 \pm 0.2$ & $2.0^{+0.4}_{-0.3}$ & 332/211 \nl
\nl T.B. &
  $4.1^{+0.6}_{-0.5}$ & \nodata &
  $0.11^{+0.05}_{-0.04}$ & $0.81^{+0.07}_{-0.06}$ & $0.51^{+0.09}_{-0.08}$ &&
 $1.3 \pm 0.2$ & $1.6 \pm 0.2$ & 366/211 \nl
\nl MEKAL &
  $3.5^{+0.4}_{-0.3}$ & \nodata &
  $0.16^{+0.06}_{-0.05}$ & $2.0 \pm 0.1$ & $1.3 \pm 0.2$ &&
  $1.5 \pm 0.2$ & $4.2 \pm 0.4$ & 351/211 \nl
\enddata
\tablecomments{ T.B. refers to thermal bremsstrahlung. Norm refers
to the normalization used for a particular model: power law ---
(photons~keV$^{-1}$~cm$^{-2}$~s$^{-1}$), thermal bremsstrahlung
--- ([$3.02 \times 10^{-15}/(4 \pi D^{2})] \int n_{e} n_{I} \:
dV$), where $n_e$ and $n_I$ are the local electron and ion
densities and $D$ is the distance, or MEKAL --- ([$10^{-15}/(4 \pi
D^{2})] \int n_{e} n_{H} \: dV$), where $n_H$ is the local
hydrogen density. All uncertainties represent the 90\% confidence
limits.} \tablenotetext{a}{Parameter for the \asca\
(GIS--2$+$GIS--3) data.} \tablenotetext{b}{Parameter for the
\rosat\ PSPC data.}
\end{deluxetable}
}

The emission from \snrG\ significantly contaminates the signal
from \axG\ and could prevent the detection of X-ray pulsations
from \psrG. To minimize this possibility, we extracted data using
combinations of four different aperture radii (ranging between
2\arcmin\ and 5\arcmin\ in increments of 1\arcmin) and twelve
different energy bands (e.g., 0.8--10.0~keV and 1.0--5.0~keV). All
resultant pulse profiles were searched for pulsations using both
the $H$-test (de Jager 1994) \nocite{dej94} and $\chi^{2}$ (Leahy
et al. 1983) \nocite{lde+83}. No pulsations were found in any data
set.

Using the method of Brazier (1994), \nocite{bra94a} we find a 95\%
confidence upper limit on the pulsed flux of 61\% for data in the
energy range 3.0--10.0~keV within a 3$'$ radius of \axG. The upper
limit is not particularly constraining, due in part to our use of
the non-standard observing mode that made pulsation searching
possible. As discussed in \S2, sacrificing Risetime information
limits {\it ASCA}'s ability to differentiate between
Earth-orbit-related background and celestial X-rays. In
particular, 2-hr binning of time series extracted from different
locations in the GIS field-of-view reveals common trends
(specifically, a rise and decline in the count rate by a factor of
$\sim$3 over a $\sim$6-hr period) that cannot be celestial. Such
behavior does not significantly affect our spectral analysis or
the observed morphology of emission, however.

\subsection{Spectral Analysis}
\label{sec:j1119_spectrum} We restricted our spectral work to only
the \asca\ GIS and \rosat\ PSPC data.\footnote{Due to the
continual radiation damage suffered by the CCDs, spectral data
from the SIS this late in {\it ASCA}'s life is highly suspect
(see, e.g., Ueda et al. 1999).} \nocite{uti+99} We defined regions
where it is appropriate to sum counts together and construct a
single spectrum for a given area. One such region is the western
side of the SNR, bright in soft X-rays. We extracted a spectrum
for both the GIS and PSPC from an ellipse $17\arcmin \times
7\arcmin$ in extent, with major axis parallel to lines of constant
right ascension and centered in the middle of the bright SNR
emission. A natural complement to this spectrum is one drawn from
the eastern side of \snrG. Here, only the GIS, with its
high-energy (i.e. $E > 2$~keV) sensitivity, can provide
spectroscopic information. In this case, the spectrum is drawn
from a crescent-shaped region, slightly larger than the radio
shell and excluding the point-source and the western side. These
regions are shown in Figure~\ref{fig:j1119_asca} ({\it top right})
and Figure~\ref{fig:j1119_rosat} ({\it right}). We also extracted
a GIS spectrum for the point source from a circular region with
radius of $2\farcm5$. The data from GIS-2 and GIS-3 were kept
separate, to avoid complications associated with summing them into
a single spectrum.

\subsubsection{\snrG}
The PSFs of both \asca\ and \rosat\ result in a contamination of
the source flux with diffuse XRB. A background spectrum must
therefore be subtracted from the data. Ideally, the background
should be taken from a nearby region that is source free and
located at the same off-axis angle, which not only accounts for
the XRB but for any local diffuse emission, always a possibility
when looking in the Galactic plane. While this task is trivial for
the PSPC and its 2\degrees\ FOV, the situation is more difficult
for the GIS. With more than 50\% of the GIS FOV unoccupied by
\snrG, it might appear this region is appropriate for extracting a
background. However, the scattering properties and broad wings of
the XRT result in reflection properties that are strongly
dependent on both incident energy and off-axis angle (see, e.g.,
Gendreau 1995). \nocite{gen95} Instead, we use another feature of
the {\tt mkgisbgd} \ftool\ to construct an accurate background. We
again refer the interested reader to Pivovaroff (2000).

Our initial attempts at spectral fitting indicated that all five
data sets could be described by a single model, as long as we
allowed for different normalization values for each telescope and
unique column densities ($N_{H}$) for each side of the SNR. In
total, there were six free parameters: two values of $N_{H}$, the
spectral characterization (e.g., temperature for a plasma model),
and three normalization values, two for the western side
(GIS-2$+$GIS-3 and PSPC) and one for the eastern side
(GIS-2$+$GIS-3). The energy range fit is restricted to those bands
where the signal is statistically significant: 0.4--2.0~keV
(PSPC), 0.7--7.0~keV (GIS-western side), 0.7--8.0~keV (GIS-eastern
side). Data were then rebinned such that each
background-subtracted energy bin had a minimum of 20 counts,
allowing us to use the $\chi^{2}$ statistic as our goodness-of-fit
estimator. All fitting was performed with {\tt XSPEC v.10.0} using
standard models.

We use a thermal bremsstrahlung model and the
MEKAL\footnote{http://heasarc.gsfc.nasa.gov/docs/journal/meka6.html.}
plasma model as representative thermal spectra and a simple
power-law to explore non-thermal models. Elemental abundances have
been frozen at the values determined by Anders \& Grevesse (1989).
\nocite{ag89} Table~\ref{tab:j1119_spectral} lists the results of
our fits, including derived parameters and their 90\% confidence
limits. The measured absorbed fluxes for the western region are
$F_{\rm \; 0.7-7.0~keV} = 2 \times 10^{-12}$~\FX\ {\it (ASCA)} and
$F_{\rm \; 0.4-2.0~keV} = 7 \times 10^{-13}$~\FX\ {\it (ROSAT)}
and for the eastern region is $F_{\rm \; 0.7-8.0~keV} = 3 \times
10^{-12}$~\FX\ {\it (ASCA)}. Although the power-law model has the
lowest $\chi^{2}$, the $F$-test (see, e.g., Bevington \& Robinson
1992) \nocite{br92} indicates that all three models describe the
data equally well. The rather large reduced $\chi^{2}$ values
($\chi_{\nu}^{2} = 1.6-1.7$) indicate a systematic discrepancy
between the data and each model.

Figure~\ref{fig:j1119_spectra} displays spectra from different
regions of \snrG. For clarity, we present the data for a given
region and instrument separately: PSPC western side ({\it top\/}),
GIS western side ({\it middle}), GIS eastern side ({\it
bottom\/}). Each panel also plots the best-fit power-law model.
(We note that due to the moderate spectral resolution of both the
GIS and PSPC, the thermal models have a similar shape to the
non-thermal one shown). The large $\chi^{2}$ is mainly
attributable to systematic error, easily seen in the residuals
below 2~keV in the spectra from the western side and above 5~keV
in the spectra from the eastern side. Attempts to improve the fit
by addition of a second spectral component do not work for two
reasons. First, while some of the excess residuals appear to be
emission-like features, there are no known lines at these
energies. Second, although the approximately 1000 counts in each
GIS spectrum and the 500 counts in the PSPC spectrum allow a
fairly tight constraint on a single model, there is not sufficient
data to fit a combination of two models.

\subsubsection{\axG} \label{sec:axG}
Since the number of point-source photons was low (173
background-subtracted counts between 0.7--5.0~keV), fitting a
spectrum is difficult. Given that it is not detected with \rosat\
nor in the soft-band of the GIS, this source must either be
intrinsically hard (e.g., temperature of several to tens of keV)
or highly absorbed. With three times more events in the
2.0--5.0~keV band than in the 0.7--2.0~keV band and very few
events above 5~keV, either scenario is plausible.

Due to the limited statistics, it is impossible to fit the
spectrum when all three parameters ($N_{H}$, $kT$ or $\Gamma$, and
normalization) are allowed to vary. Instead, we only required 10
background-subtracted counts per energy bin and fixed the column
density at $N_{H} = 1.5 \times 10^{22}$~cm$^{-2}$, the value
determined using the MEKAL model for the eastern side of \snrG.
While these modifications to the fitting scheme, compared to
those used for \snrG, introduce additional uncertainty, they do
allow at least a coarse characterization of the spectral nature of
the point source.

For a power-law model, the photon index is $\Gamma =
1.4^{+1.0}_{-1.2}$ and the normalization is $(8 \pm 8) \times
10^{-5}$. For a thermal bremsstrahlung model, the temperature is
$kT = 13^{+\infty}_{-11}$~keV and the normalization is
$(1^{+0.9}_{-0.3}) \times 10^{-5}$. (Here, the upper limit on $kT$
reflects the lack of {\it ASCA}'s response above 10~keV. Refer to
Table~\ref{tab:j1119_spectral} for the normalization units.) The
measured absorbed flux is $F_{\rm \; 0.7-5.0~keV} = 2 \times
10^{-13}$~\FX.

\bigskip
\bigskip
\epsfxsize=8truecm \epsfbox{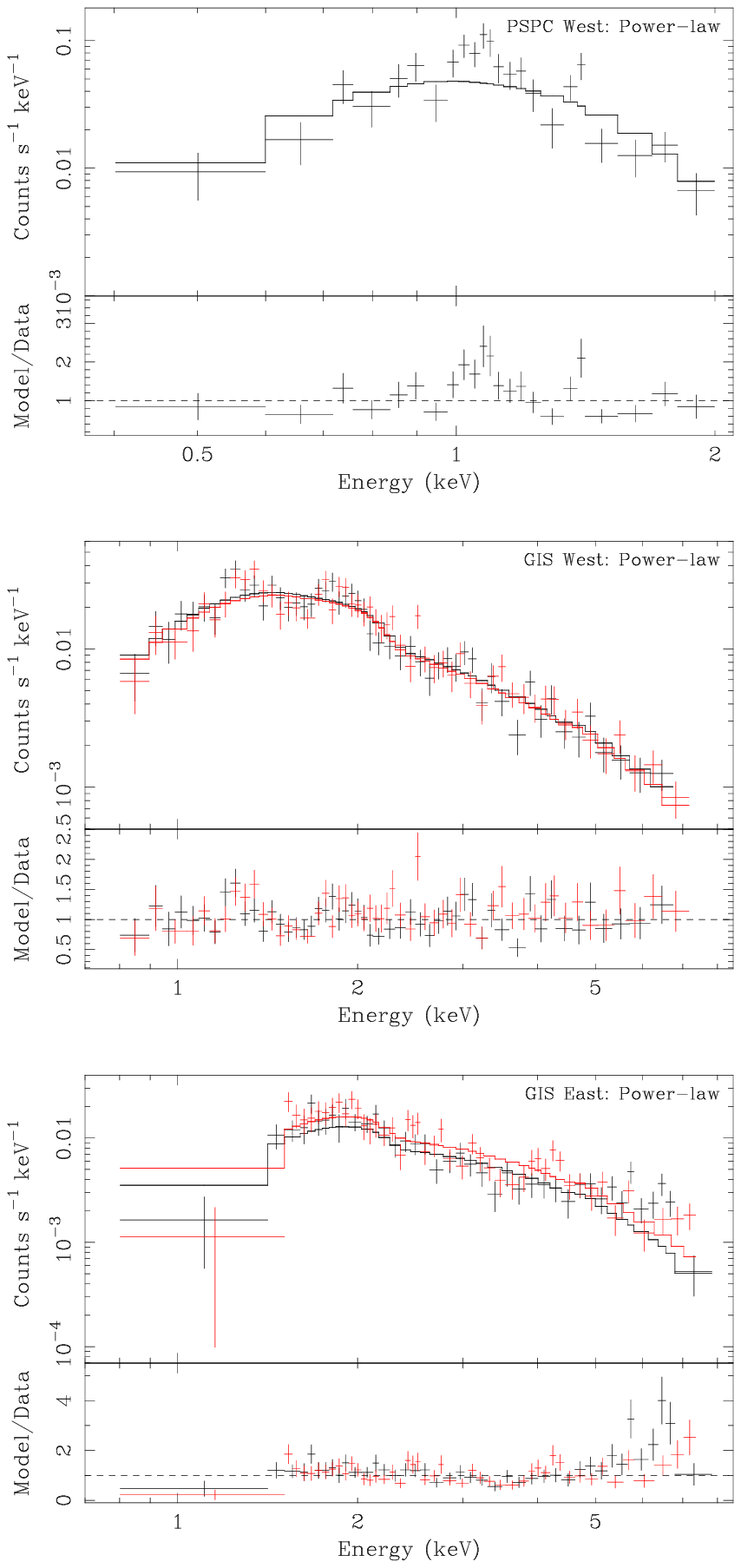}
\figcaption[mjp3.eps]
{\label{fig:j1119_spectra} \asca\ GIS and \rosat\ PSPC spectra of
\snrG. The PSPC ({\it top\/}) and GIS ({\it middle}) spectra from
the western side of the SNR are very soft compared to the highly
absorbed GIS spectrum ({\it bottom\/}) of the eastern side. The
solid lines represent the best-fit power-law model to the data. In
the bottom two panels, data from GIS--2 are in black, data from
GIS--3 in red.}
\bigskip

\section{Discussion}
\subsection{General Properties of \snrG}
\label{sec:j1119_general} One of the main arguments supporting the
interpretation of the extended X-ray emission as a SNR is its
correlation with the morphology of the radio shell, recently shown
to have radio spectral properties consistent with those of other
known SNRs (Crawford et al. 2001). The presence of \psrG\ at the
center of the radio and X-ray emission strengthens this reasoning
and provides a way to test the consistency of the claimed
association. If \psrG\ and \snrG\ are the end-products of the same
supernova, their ages should be the same and the properties of the
SNR should be those of a young remnant.

The age of \psrG\ as estimated from its spin parameters is
$1700\pm100$\,yr (\cite{ckl+00}). This assumes that the initial
spin period $P_0$ of the neutron star was very small compared to
the present one; if, however, $P_0$ were larger, the age would be
smaller. In any case, as measured from the spin parameters, $\sim
1800$\,yr represents a good upper limit on the age, as the braking
index has been measured (Table~1).

A distance estimate can be obtained from the Taylor \& Cordes
(1993) DM--distance relationship. \nocite{tc93} The DM of \psrG\
implies a distance of more than 30~kpc, well outside the
Galaxy. This implausibly large value is easily understood, as the
Galactic longitude of \psrG\ ($l = 292 \degrees$) is nearly
tangential to the Carina spiral arm, intersecting it at distances
of 2.4 and 8.0~kpc. The best explanation for the large DM is that
there is clumping of dense, dispersive material in the direction
of the pulsar. The Taylor \& Cordes model, which lacks fine
structure (i.e. small-scale clumps), likely underestimates the
electron density and hence overestimate the distance for this
line-of-sight. A reasonable upper limit on the distance can be
obtained by assuming that \psrG\ lies no further than the second
intersection point, or 8~kpc. We also note that in this
direction, the edge of the Galaxy only extends $\sim$10~kpc from
the Sun (Georgelin \& Georgelin 1976; Taylor \& Cordes 1993),
\nocite{gg76} \nocite{tc93} resulting in a maximum error on the
distance upper limit of 25\%. We adopt a compromise distance of
5~kpc for the calculations below.

The diameter of \snrG\ extends $\sim$15\arcmin\ in radio and
$\sim$17\arcmin\ in X-rays. The slightly larger extent in X-rays
may be real or an artifact of the larger PSFs of \asca\ and
\rosat, compared to the ATCA beamsize. We adopt an angular size of
$(15 \pm 2$)\arcmin\ to encompass both measurements. Taken with
the age and distance estimates, we calculate a mean expansion
velocity of $v = (6.2 \pm 0.9) D_{5} \times 10^{3} \; {\rm km} \;
{\rm s}^{-1}$, where $D_{5}$ is the distance to the pulsar
parameterized in units of 5~kpc. This velocity is consistent with
that of a 1700-yr-old SNR still evolving in the free expansion
phase. It implies a kinetic energy for the initial explosion of
$E_{51} = (0.4 \pm 0.1) M_{\rm ej} \: D_{5}^{2}$, where $E_{51}$
is the explosion energy in units of $10^{51}$ ergs and $M_{\rm
ej}$ is the ejected mass in solar masses.

\snrG\ may also be in the Sedov-Taylor (ST) phase. For adiabatic
expansion, $r_{\rm SNR} = 1.15 (Et^{2}/\rho)^{1/5}$ (Taylor 1950;
Sedov 1959), where $r_{\rm SNR}$ is the linear radius of the SNR,
$E$ is the explosion energy, $t$ is the SNR's age, and $\rho$ is
the mass density of the ambient medium into which the SNR is
expanding. \nocite{tay50,sed59} Using the measured size of the SNR
and recasting in terms of the ambient particle density $n$,
$E_{51}/n = (12 \pm 8) D_{5}^{5}$. Remnants only enter ST
evolution after sweeping up $\sim 20M_{\rm ej}$ (Fabian,
Brinkmann, \& Stewart 1983; Dohm-Palmer \& Jones 1996).
\nocite{fbs83,dj96} We estimate the material swept up by the SNR
by assuming a constant density $\rho = n m_{\rm H}$ inside the
volume occupied by \snrG; this requires $n > (0.16 \pm 0.08)
M_{\rm ej} \: D_{5}^{-3} \; {\rm cm}^{-3}$. Therefore, $E_{51} >
(2.0 \pm 1.6) M_{\rm ej} D_{5}^{2}$.

Thus, for either free or adiabatic expansion, the implied ratio
$E_{51}/M_{\rm ej}$ is much greater than those found for most
young Galactic SNRs (see, e.g., Smith 1988). \nocite{smi88c} We
note that the only other well documented large value for
$E_{51}/M_{\rm ej}$ is for G320.4$-$1.2, the SNR associated with
PSR~B1509$-$58 (Gaensler et al. 1999 and references therein).
\nocite{gbm+99b} This is intriguing as \psrG\ and PSR~B1509$-$58
are very similar (see \S1, Table~1); it raises the possibility
that their large magnetic fields are related to a common
evolutionary scenario. For example, Gaensler et al. (1999) suggest
that the progenitor of PSR~B1509$-$58 was a massive star that
later evolved into a helium star before it underwent a supernova.

\subsection{X-ray Properties of \snrG}
\subsubsection{Spectrum}
In principle, one of the most surprising aspects of the spectrum
is the lack of obvious line features commonly seen in young SNRs
like Cas~A (e.g., Hughes et al. 2000), Tycho (Hwang \& Gotthelf
1997), MSH 15$-$5{\it 2} (e.g., Tamura et al. 1996), and Pup~A
(Winkler et al. 1981; Berthiaume et al. 1994). \nocite{hrbs00}
\nocite{hg97} \nocite{tkyb96} \nocite{wcc+81} \nocite{bbgn94}
However, from the MEKAL fits, it is clear that after the line-rich
spectra are folded through the GIS response, the only lines that
should be visible are possibly the Fe K species around 6.7~keV. In
fact, the residuals above 6~keV are smaller for the MEKAL model
than for the power-law model, indicating that these lines may
indeed be present.

{\scriptsize
\begin{deluxetable}{l c c}
\tablewidth{4in} \tablecaption{X-ray temperature dependence on
elemental abundances \label{tab:j1119_abund}} \tablehead{
\colhead{$kT$} & \colhead{Abundance} & \colhead{} \\
\colhead{(keV)} & \colhead{factor\tablenotemark{a}} &
\colhead{$\chi^{2}$/dof} } \startdata $3.5^{+0.4}_{-0.3}$ & 1.0 &
351/211 \nl $3.4 \pm 0.3$ & 1.5 &  376/211 \nl $3.1 \pm 0.3$ & 2.0
& 410/211 \nl $2.8 \pm 0.2$ & 3.0 & 479/211 \nl
\enddata
\tablenotetext{a}{Here, only the most common heavy elements (Si,
S, and Fe) have had their abundances, as determined by Anders \&
Grevesse (1989), \nocite{ag89} multiplied by this factor.}
\end{deluxetable}
}

One difficulty with either of the thermal models is the large
derived temperature of $kT \approx 4$~keV. Typically, even young
remnants with ages less than 1000~yr have temperatures closer to
2~keV (Koyama et al. 1996; Sakano et al. 1999). \nocite{kms+96}
\nocite{sym+99} One plausible explanation for this is the use of
a simple plasma model assuming solar elemental abundances. In
most cases, SNR spectra with sufficiently high counting statistics
(a minimum of several thousand source counts) or high energy
resolution ($E/\Delta E$ of several hundred, as in the case of the
Focal Plane Crystal Spectrometer on \einstein) require non-solar
abundances and at least one non-equilibrium ionization (NEI) model
to properly describe the data (e.g., Hughes et al. 2000; Winkler
et al. 1981; Hayashi et al. 1994). \nocite{hrbs00} \nocite{wcc+81}
\nocite{hko+94}

While the lack of any obvious line features prevents us from
attempting to fit more realistic models, we explored this
possibility by adjusting the abundances of the three metals with
the largest number densities relative to H, namely Si, S, and Fe.
Table~\ref{tab:j1119_abund} lists the best-fit MEKAL temperatures
obtained when the abundances have been multiplied by factors of
1.5, 2.0 and 3.0. As the amount of metals is increased, the
temperature monotonically drops from $3.5^{+0.4}_{-0.3}$~keV to
$2.8 \pm 0.2$~keV. The goodness of fit ($\chi^{2}$) also grows, as
do the systematic residuals, indicating that abundances factors of
five to ten greater than solar are ruled out. However, it is quite
realistic to expect that the use of a NEI model in conjunction
with modestly-enriched abundances would result in a goodness of
fit comparable to that of the power-law or MEKAL model.

The spectral fitting we performed also allows the intriguing
possibility that the emission from \snrG\ is non-thermal in
origin. It is well-established that most young SNRs have a strong
non-thermal component (see Allen, Gotthelf, \& Petre 1999 for a
recent review). \nocite{agp99} In the most commonly accepted
scenario, electrons are accelerated by the remnant's shock wave to
energies of $\sim$1~TeV and emit high-energy radiation via the
synchrotron mechanism (e.g., Reynolds 1998). \nocite{rey98}
Usually, though, thermal X-rays are also present in the SNR. One
notable exception is SNR~G347.3$-$0.5, which shows no measurable
thermal emission down to very low limits (Slane et al. 1999).
\nocite{sgd+99} The photon index $\Gamma$ measured for \snrG\
($2.3 \pm 0.1$) agrees well with those in the range of values
reported for different regions of G347.3$-$0.5 (2.2, 2.4, and
2.4). This spectrum is distinctly harder than those of SNRs that
exhibit both thermal and non-thermal emission, like Cas~A ($\Gamma
= 3.0 \pm 0.2$), SN~1006 ($3.0 \pm 0.2$), Kepler ($3.0 \pm 0.2$),
Tycho ($3.2 \pm 0.1$), and RCW~86 ($3.3 \pm 0.2$) (Koyama et al.
1995; Allen et al. 1997; Allen et al. 1999). \nocite{agp99}
\nocite{akg+97} \nocite{kkm+97}

Although the spectral similarities support the idea that \snrG\
may belong to a class of non-thermal remnants typified by
G347.3$-$0.5, Slane et al. (1999) show that the properties of
G347.3$-$0.5 can reasonably be explained if the remnant is in a
well-advanced Sedov evolutionary phase and has an age between
19--41~kyr. This is in contrast to \snrG, which is extremely young
and is (possibly) just entering the Sedov phase. Ultimately, the
nature of \snrG, be it a typical young thermal SNR or a more
exotic manifestation of the SNR phenomenon, will only be decided
with additional observations.

\subsubsection{Help from Nearby Objects}
Another important aspect of the spectrum that requires explanation
is the lack of soft X-rays from the eastern side of \snrG\ and
the rather uniformly filled morphology at high energies (compare
Figures~\ref{fig:j1119_asca} and \ref{fig:j1119_rosat}). The
absence manifests itself via absorption of emission below
$\sim$1.5~keV (Figure~\ref{fig:j1119_spectra} [{\it bottom
row\/}]). Using SIMBAD, we looked for objects in the vicinity of
\psrG\ that might account for the absorption. A likely candidate
is Dark Cloud DC~292.3$-$0.4, catalogued by Hartley et al. (1986)
during a systematic search of ESO/SERC Southern J survey plates
for optically identified dark clouds. Their work is an extension
of the work by Lynds (1962) to declinations south of --35\degrees.
\nocite{hms+86} \nocite{lyn62} Hartley et al. (1986) approximate
the shape and size of each cloud with an ellipse and use three
classes to characterize the density of each cloud. (The reported
dimensions do not necessarily reflect the shape of the cloud
[e.g., if the cloud is elongated or curved], but do give an
accurate estimate of its total area.)

DC~292.3$-$0.4 is described by a circle with diameter 16\arcmin.
Figure~\ref{fig:j1119_clouds} shows the dark cloud, represented by
a hatched-circle, plotted on the hard-band PSPC image (for
brevity, we do not show the similar GIS image). DC~292.3$-$0.4
appears to be located at a position capable of obscuring the
eastern side of \snrG, and given that the cloud is certainly not
spherical, it seems quite plausible that substructure (e.g., a
finger or wisp) extending across the SNR absorbs the soft X-ray
emission.

\bigskip
\epsfxsize=8truecm \epsfbox{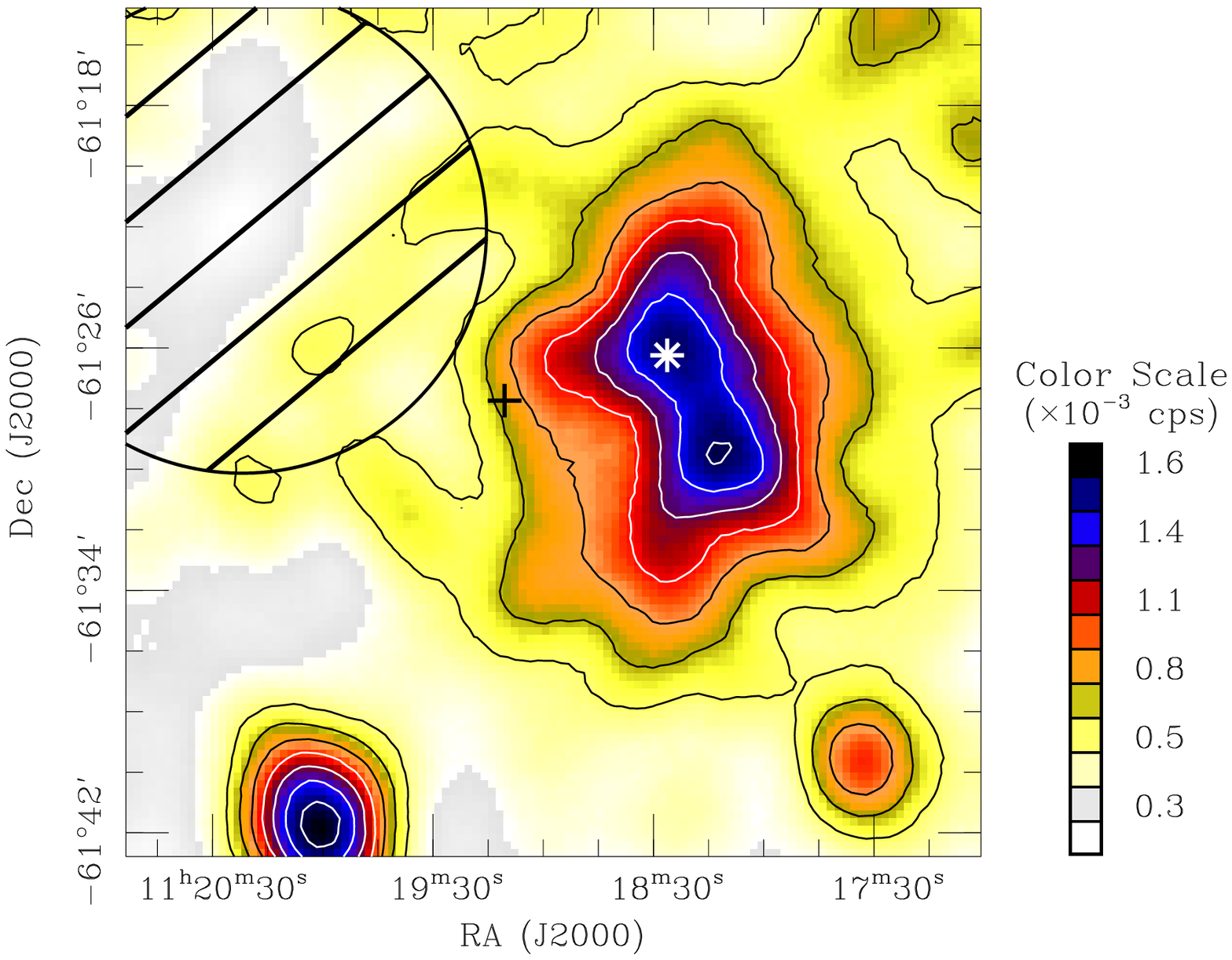} \figcaption[mjp4.eps]
{\label{fig:j1119_clouds} Hard-band PSPC image  of \snrG.  The
large hatched circle represents the approximate shape of dark
cloud DC~292.3$-$0.4, which we suggest accounts for a large part
of the absorption of soft X-rays from the eastern side of the SNR.
The pulsar location is marked by a cross.  The asterisk marks the
location of HD~306313, a B9 star that is positionally coincident
with enhancements in emission from the western side of the remnant
in both detectors.  Contours span between 35\% -- 95\% of the
maximum flux in increments of 10\%.}
\bigskip

A more quantitative check is to see if the cloud can account for
the difference in column densities ($N_{H} \simeq 1.3 \times
10^{22}$~cm$^{-2}$) between the two sides of the SNR. The cloud's
density (class B) roughly equals the Lynds designation of opacity
class (OC) 4 or 5, which, using the calibrated relationship of
Feitzinger \& St\"{u}we (1986), $A_{V} = 0.70 \; {\rm OC} +
0.5$~mag, gives an extinction $A_{V} = 3.3 -4.0$ from
DC~292.3$-$0.4. \nocite{fs86} The corresponding column density
$N_{H}$ can be estimated using $N_{H} = 1.7 \times 10^{21} \;
A_{V} \; {\rm cm}^{-2} \; {\rm mag}^{-1}$, derived from $\langle N
{\rm (H}${\sc i)}$/E(B-V) \rangle = 5.2 \times
10^{21}$~cm$^{-2}$~mag$^{-1}$ (Shull \& van Steenberg 1985)
\nocite{sv85} and $A(V)/E(B-V) = 3.1$ (Cardelli, Clayton, \&
Mathis 1989). \nocite{ccm89} Thus, we expect DC~292.3$-$0.4 to
contribute $(6-7) \times 10^{21}$~cm$^{-2}$ to the eastern side of
\snrG, or roughly half of the additional amount of $N_{H}$ present
on this side of the SNR.

The presence of DC~292.3$-$0.4 also offers the chance to probe the
distance to PSR J1119$-$6127. Recently, Otrupcek, Hartley \& Wang
(2000) \nocite{ohw00} observed the 115 GHz (J=1$-$0) transition of
CO towards the center of each cloud in the Hartley et al. catalog.
Two features with line-of-sight velocities of --12.6 km~s$^{-1}$
and 1.6 km~s$^{-1}$ were detected. Unfortunately, with no way to
discern which feature corresponds to the cloud, and as each
feature implies two possible distance values, these CO
measurements cannot provide a constraining lower limit for the
pulsar's distance. The presence of two features is encouraging,
however, as it suggests an additional cloud is present along the
line of sight and also contributes to the absorption of soft
X-rays from the eastern side of \snrG.

Figure~\ref{fig:j1119_clouds} also shows the location of HD~306313
(marked with an asterisk), a B9 star with apparent magnitude 11.6.
Until recently, late B stars were not thought to exhibit
observable high energy emission. However, analysis of the \rosat\
all-sky survey revealed that these stellar types can in fact be
X-ray bright (Bergh\"{o}fer \& Schmitt 1994; Bergh\"{o}fer,
Schmitt, \& Cassinelli 1996). \nocite{bs94,bsc96} While only 10\%
of B9 stars emit X-rays (Bergh\"{o}fer et al. 1997),
\nocite{bsdc97} this offers a possible explanation for the flux
enhancements in both the PSPC and GIS images near the star. The
USNO-A2.0 catalog of stars gives a blue magnitude of 12.7 and a
red magnitude of 11.6 for HD~306313. The USNO calibration
algorithms\footnote {http://www.nofs.navy.mil.} allow us to
convert to standard $B$ and $V$ colors, and assuming a magnitude
uncertainty $\sigma = 0.25$ (Grazian et al. 2000) \nocite{gcd+00}
and correcting for the intrinsic color of a B9 star, we calculate
color excesses $\langle E(B-V) \rangle$ for several stellar
classes (i.e. {\sc i, iii, v}). Finally, adopting the absolute
magnitudes measured by Jaschek \& G\'{o}mez (1998) \nocite{jg98}
for B9 stars and the reddening law used above, we derive distances
of $250 \pm 80$~pc (B9~{\sc v}), $550 \pm 165$~pc (B9~{\sc iii}),
and $6.3 \pm 3.5$~kpc (B9~{\sc i}).

The PSPC flux (see below) from the entire western region totals $\sim 7
\times 10^{-13}$ \FX. We estimate that the star would only need to
have $1-10$\% of this flux to be observable. The X-ray luminosity in
the 0.1--2.4~keV band for B9 stars ranges between ${\rm log}\: (L_{x})
= 28.5 -31.0$ (Bergh\"{o}fer et al. 1997). While the low flux
precludes emission from a distant (i.e. more than a few kpc)
supergiant, a main sequence or giant star, with maximum unabsorbed
fluxes of $1.3 \times 10^{-12}$ and $2.8 \times 10^{-13}$~\FX, could
easily result in the bright feature visible in the hard-band PSPC and
soft-band GIS data. Moreover, stellar emission contamination of the
SNR spectrum would also explain the systematic residuals seen below
2~keV. In a detailed spectral study of A0--F6 stars, Panzera et al.
(1999) \nocite{ptpa99} show that these stars are best described by a
combination of two Raymond-Smith plasma models with average
temperatures of $\langle kT \rangle \sim 0.7$~keV and $\langle kT
\rangle \sim 0.2$~keV. We tried adding a Raymond-Smith component to
our best-fit models, but for the reasons stated in
$\S$\ref{sec:j1119_spectrum}, these attempts were unsuccessful.

We conclude with an intriguing possibility to be pursued in future
observations. If HD~306313 is confirmed as an X-ray source, with
sufficient data its spectral properties can be determined. By
comparing its column density with that measured for the western
side of \snrG, an upper or lower limit to the pulsar/remnant
system can be obtained.

\subsubsection{X-ray Luminosity}
Table~\ref{tab:j1119_flux} presents the flux measured from each
side of \snrG\ for each  spectral model and each instrument.
Luminosities have been calculated assuming a distance of 5~kpc and
correcting for interstellar absorption. Formal errors on $L_{x}$
are $\sim$5\%, and regardless of the spectral model, the
luminosities for a given region and instrument are within 20\% of
one another, guaranteeing a reliable measurement of the total
luminosity from \snrG. While the \rosat-measured luminosities are
consistently lower than those of \asca, this is easily understood
given the uncertainties in calibration and differences in
telescope sensitivities. The ratio between eastern and western
side luminosities from \asca\ is $2.0 \pm 0.2$, in excellent
agreement with the 2.1 ratio of geometric areas of each region.
The total 0.5--10.0~keV luminosity from \snrG\ is $(3-4) \times
10^{35}$~\LX, after correcting for absorption, and assuming a
distance of 5\,kpc.

{\scriptsize
\begin{deluxetable}{l  cc c cc  c  cc}
\tablewidth{6in} \tablecaption{X-ray flux and luminosity for SNR
G292.2$-$0.5 \label{tab:j1119_flux}} \tablehead{ \colhead{} &
\multicolumn{5}{c}{Western Side} & \colhead{} &
\multicolumn{2}{c}{Eastern Side} \\ \cline{2-6} \cline{8-9}
\colhead{} & \multicolumn{2}{c}{ASCA} &  \colhead{}&
\multicolumn{2}{c}{ROSAT} & \colhead{}&  \multicolumn{2}{c}{ASCA}
\\ \cline{2-3} \cline{5-6}  \cline{8-9} \colhead{Spectral} &
\colhead{$F_{\rm 0.7-7.0}$} & \colhead{$L_x$} & \colhead{}&
\colhead{$F_{\rm 0.4-2.0}$} & \colhead{$L_x$} & \colhead{} &
\colhead{$F_{\rm 0.7-8.0}$} & \colhead{$L_x$}  \\
\colhead{model} & \colhead{($10^{-12}$)} & \colhead{($10^{35}$)} &
\colhead{}& \colhead{($10^{-12}$)} & \colhead{($10^{35}$)} &
\colhead{} & \colhead{($10^{-12}$)} & \colhead{($10^{35}$)} }
\startdata Power law &
  2.4 & $1.1 \pm 0.1$ &&
  0.67 & $0.70^{+0.12}_{-0.08}$ & &
  3.3 & $2.34^{+0.12}_{-0.08}$ \nl
\nl Thermal brems.& 2.4 &   $0.94^{+0.07}_{-0.04}$ && 0.68 &
$0.58^{+0.12}_{-0.08}$ & & 3.1 &   $1.79\pm 0.08$ \nl \nl MEKAL&
2.4 & $0.94 \pm 0.04$ && 0.66 & $0.62 \pm 0.08$ & & 3.3 & $1.99
\pm 0.04$ \nl
\enddata
\tablecomments{All fluxes, in units of (\FX), refer to the
measured absorbed flux for the given energy band, which is in keV.
All luminosities, in units of (\LX), are for the 0.5--10.0~keV
passband. They have been corrected for absorption and assume a
distance of 5~kpc. All uncertainties represent the 90\% confidence
limits.} 
\end{deluxetable}
}

\subsection{\axG}

Figure~\ref{fig:j1119_iras} shows a close-up of the GIS hard-band
field surrounding \psrG. The radio position of \psrG\ is marked
by a cross, while an ellipse shows the 95\% confidence uncertainty
position of the unidentified infrared point-source \irasG. The
position of the IRAS source (J2000), ${\rm R.A.} = 11^{\rm
h}~19^{\rm m}~07\fs05$, ${\rm Decl.} =
-61\degrees~27^{\prime}~26\farcs3$, is offset 69\arcsec\ from
\axG. (Recall that \psrG\ is offset 87\arcsec\ from \axG.) While
the position of \irasG\ has a large uncertainty and is slightly
closer to the \asca\ source than \psrG, the offsets are too large
to claim that \axG\ is the X-ray counterpart of the pulsar or IRAS
source based solely on positional coincidence. We now consider
additional evidence of relevance for both possibilities.

\bigskip
\epsfxsize=8truecm \epsfbox{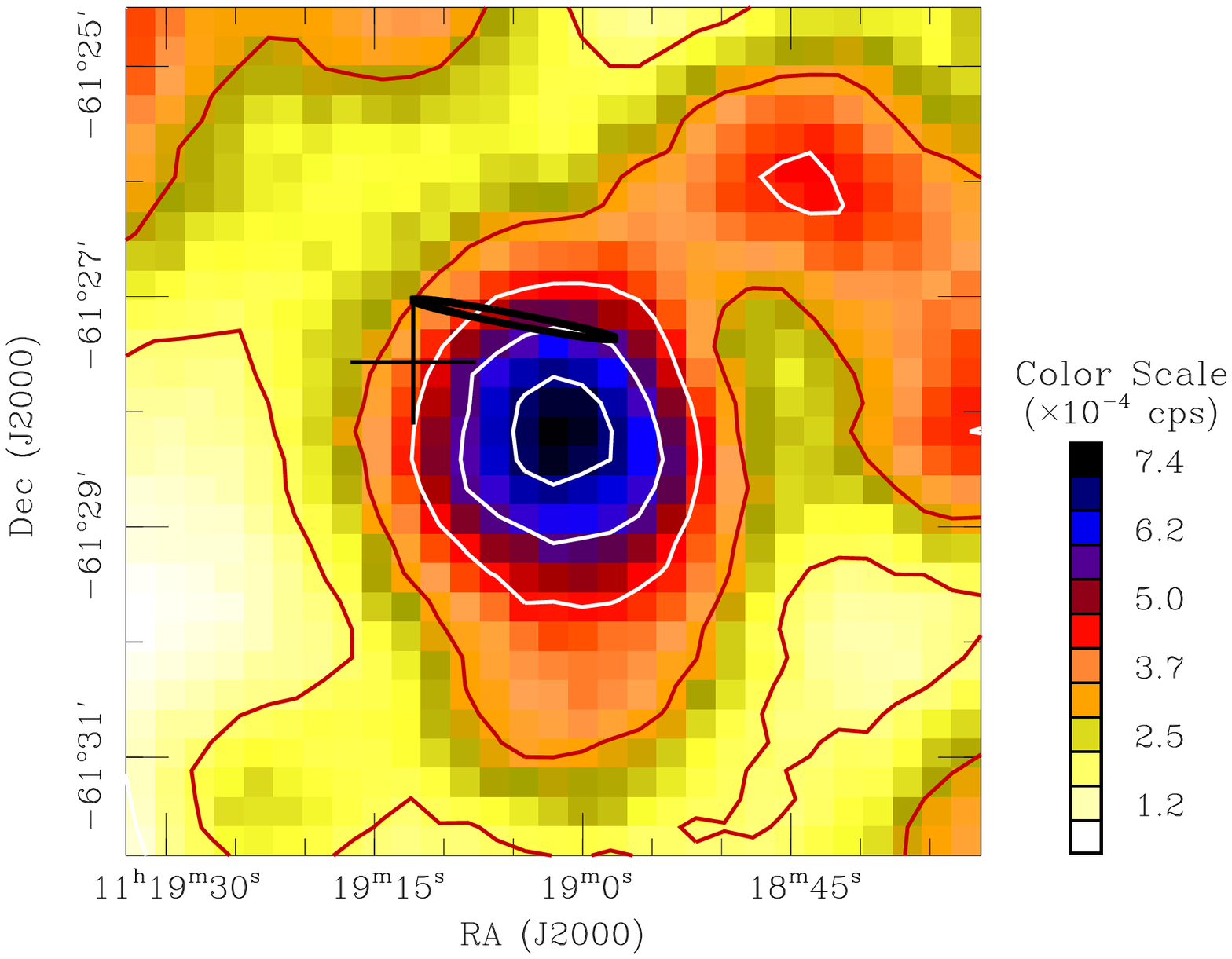} \figcaption[mjp5.eps]
{\label{fig:j1119_iras} Hard-band \asca\ image of the immediate
region around \axG.  A dark cross marks the location of \psrG\
(uncertainty much smaller than cross size), while a dark ellipse
(semi-major axis $53\arcsec$, semi-minor axis $3\arcsec$) marks
the 95\% confidence position of \irasG.  Contours correspond to
$35-95$\% of the maximum flux in increments of 10\%.}
\bigskip

\subsubsection{X-ray Luminosity}
The small number of counts from the point source precludes
determining the nature of the underlying emission mechanism (e.g.,
thermal or non-thermal). However, by deriving the luminosity
implied for various spectral models, it is possible to check the
consistency of the assumption. All the luminosities reported
below assume a distance of 5~kpc and have been corrected for the
effects of interstellar absorption. In the case when both the
photon index and normalization were allowed to vary, the implied
luminosity in the 0.1--2.4~keV band is $\left( 0.8^{+2.7}_{-0.8}
\right) \times 10^{33}$~\LX, while in the 0.5--10.0~keV band it is
$\left( 2^{+25}_{-2} \right) \times 10^{33}$~\LX. The huge spread
is due to the large uncertainty in the photon index (recall
$\Gamma = 0.2-2.4$; \S\ref{sec:axG}). When the photon index is
fixed at the usual value for young pulsars ($\Gamma = 2$), the
luminosity range narrows to $\left( 2.0 \pm 0.8 \right) \times
10^{33}$~\LX\ for both the 0.1--2.4 and 0.5--10.0~keV bands. For
the thermal bremsstrahlung model, the formal confidence limit on
the temperature is $kT = 13^{+\infty}_{-11}$~keV. If we estimate
the upper limit at $kT = 20$~keV, the luminosity in the
0.1--2.4~keV band is $\left( 0.8 \pm 0.4 \right) \times
10^{33}$~\LX, while in the 0.5--10.0~keV band it is $\left(
1.6^{+1.6}_{-1.2} \right) \times 10^{33}$~\LX. Thus, for either
the thermal or non-thermal model, the luminosity for \axG\ in
either commonly reported band (0.1--2.4 or 0.5--10.0~keV) is $L_x
\sim (0.4-3.2) \times 10^{33}$~\LX.

\subsubsection{An Unresolved Synchrotron Nebula?}
If \axG\ truly has a non-thermal spectrum described by a
relatively flat power law (i.e. $\Gamma \lesssim 2$), the implied
X-ray luminosity is consistent with the interpretation that it is
the X-ray counterpart to \psrG. First, we consider  the luminosity
with a fixed photon index $\Gamma = 2$. The conversion efficiency
of $\dot{E}$ into $L_{x}$ is $\epsilon \equiv (L_{x}/\dot{E}) =
(0.8 \pm 0.4) \times 10^{-3}$ for both the \rosat\ (0.1--2.4~keV)
and \einstein\ (0.2--4.0~keV) bands. These values are close to
those predicted by the $L_{x}$--$\dot{E}$ relationships of Becker
\& Tr\"{u}mper (1997 [$\epsilon = 1 \times 10^{-3}$]) and Seward
\& Wang (1988 [$\epsilon = 4 \times 10^{-3}$]). \nocite{bt97,sw88}
However, this apparent agreement must be viewed cautiously for
three reasons. First, as shown by Pivovaroff et al. (2000a),
\nocite{pkg00} both of these empirical relationships have inherent
scatter of at least a factor of four. Second, we have assumed a
distance of 5~kpc for \psrG; a reasonable range of uncertainty in
this distance (\S4.1) permits a range in $L_x$, and hence in
$\epsilon$, of a factor of ten. Finally, we also note that the
uncertainty in $\epsilon$ greatly increases when we consider the
luminosities derived from the spectral fits where both the
normalization and spectral index were free parameters. More than
the specific value of $\epsilon$ or whether it agrees well with a
particular $L_{x}-\dot{E}$ relationship is the order of magnitude
value: converting one part in a thousand of $\dot{E}$ into X-rays
is consistent with the majority of X-ray-detected rotation-powered
pulsars. If the pulsar is powering the observed high-energy
emission, the lack of pulsations coupled with the point-like
nature of the source argues that \axG\ is an unresolved
synchrotron nebula powered by \psrG. This would be analogous to
the X-ray emission observed by \asca\ from \psra\ (Pivovaroff et
al. 2000a), a pulsar with an $\dot E$ similar to that of \psrG\
but with $\tau_c = 20$\,kyr.

\subsubsection{A Precursor LMXB?}
If the \asca\ source has a hard ($kT > 2$~keV) thermal spectrum,
no theoretical model nor previous observational evidence supports
interpreting \axG\ as the counterpart to \psrG. Given the
point-like nature of the \asca\ source, the emission from \axG\
could result from accretion onto a compact object. This scenario
is strengthened by the nearby presence of \irasG, an infrared
point source. Recently, two different collaborations have studied
this object because of its spatial coincidence with a known S
star, a red giant similar to M-class giants with prominent ZrO
bands.

Chen, Gao \& Jorissen (1995) \nocite{cgj95} claim that \irasG\ is
actually a blend of three sources. Lloyd Evans \& Little-Marenin
(1999) \nocite{lelm99} discovered two ``very red'' objects at the
IRAS position, although their observations resolved only a single
object at the telescope. They re-classify the (possibly) composite
spectrum as M3. Although isolated late-type stars can  emit
X-rays, their spectra are very soft ($kT < 0.5$~keV; H\"{u}nsch et
al. 1998 and references therein) \nocite{hssz98} and cannot
explain the emission from \axG. Late-type giants, including M and
S stars, in binary systems with white-dwarfs can have slightly
harder spectra, with $kT$ approaching $\sim$1~keV (Jorissen et al.
1996; H\"{u}nsch et al. 1998), \nocite{jsc+96} although such a
system would still not be hard enough to explain the \asca\
source. Even if this star had unprecedented hard emission similar
to \axG, it would also need a ratio of X-ray flux to bolometric
flux several orders of magnitude higher than all other known
X-ray-bright M stars (H\"{u}nsch et al. 1998).

A more plausible explanation is provided by analogy with the hard
X-ray emitter 2A~1704$+$241 (4U~1700$+$24). This X-ray source was
first identified in the {\it Ariel V} 2A catalog (Cooke et al.
1978) and reconfirmed in the fourth {\it Uhuru} catalog (Forman et
al. 1978). \nocite{crm+78,fjc+78} Using data from \einstein\ and
{\it HEAO--1}, Garcia et al. (1983) found the spectrum well
described by a highly absorbed ($N_{H} \sim 10^{22}$~cm$^{-2}$),
hard ($kT = 15$~keV) thermal bremsstrahlung model and a
2.0--11.0~keV luminosity between $10^{33}-10^{34}$~\LX. They also
identified the M3 giant HD~154791 \nocite{gbd+83} as its optical
counterpart.

More recently, Gaudenzi \& Polcaro (1999) \nocite{gp99} performed
detailed optical spectroscopy of this system and hypothesized that
the observed X-ray emission is powered by accretion onto a neutron
star. Moreover, they claim that this system is in the process of
evolving into a normal LMXB. Dal Fiume et al. (2000)
\nocite{dmb+00} present recent \asca\ and \bepposax\ observations
of 4U~1700$+$24. In agreement with the results of Garcia et al.
(1983), they find that the emission is best described with a
thermal bremsstrahlung model, although with lower temperatures of
$kT = 6.3$~keV and $kT = 3.6$~keV for \asca\ and \bepposax,
respectively. They also show that during both the \asca\ and
\bepposax\ observations, each of which spans 40~ks, the source
experienced variability by a factor of two. Equally intriguing,
the time-averaged luminosities derived for the \asca\ and
\bepposax\ observations varied by nearly a factor of three. In
contrast to Gaudenzi \& Polcaro (1999), Dal Fiume et al. (2000)
claim that the X-ray emission is powered by accretion from the M3
giant onto a white dwarf. Whatever the nature of compact object, a
similar scenario of accretion from \irasG\ onto a neutron star or
white dwarf could explain both the luminosity and spectrum of
\axG. In this case, the X-ray binary would be completely unrelated
to \psrG. We note that because of the absence of the Risetime
information for our {\it ASCA} data mode (see \S2) and the
resulting poor background subtraction (see \S3.2), we cannot rule
out variability in \axG.

\section{Conclusions}

We have detected X-ray emission from the direction of the young
radio pulsar \psrG\ in observations with the {\it ASCA} and {\it
ROSAT} satellites. X-ray emission is detected from a nearly
circular region of diameter $\sim 17'$ that roughly matches the
morphology of the shell-type SNR observed at radio wavelengths. We
identify this as the X-ray-bright SNR~\snrG. Without further
observations it is not possible to characterize the spectral
properties of this SNR, with either thermal or non-thermal
emission possible. The total 0.5--10.0~keV luminosity from \snrG\
is $(3-4) \times 10^{35}$~\LX, after correcting for absorption,
and assuming a distance of 5\,kpc.

The radio pulsar is located near the center of the shell of radio
and X-ray emission, and we have detected a hard point-like X-ray
source about $1\farcm5$ southwest of the pulsar with the {\it
ASCA} GIS instrument. While this offset is larger than expected
from the fitting procedure and {\it ASCA} pointing uncertainty, it
is not impossibly so, and we consider the possible association of
the X-ray point source \axG\ with \psrG. No pulsations are
detected from this source, with a 95\% confidence upper limit of
61\% for a sinusoidal profile. It is not possible to fit a unique
spectral model to the point source, due to limited counting
statistics, however the data are well described by a power-law
model with photon index $\Gamma \approx 1-2$. For either thermal
or non-thermal models, the luminosity for \axG\ in either the
0.1--2.4 or 0.5--10.0\,keV bands is $L_x \sim (0.4-3.2) \times
10^{33}$~\LX, assuming a distance of 5\,kpc. This luminosity is
(0.2--1.2)$\times 10^{-3} \dot E$, where $\dot E = 2.3\times
10^{36}$~\LX\ is the pulsar spin-down luminosity. This, of course,
is the maximum luminosity being produced in X-rays by the pulsar
plus plerion combination: if \axG\ is not associated with \psrG,
$L_x$ is smaller, i.e., $L_x/\dot E \simlt 0.1\%$. For comparison,
PSR~B1509$-$58, the radio pulsar ``most similar'' to \psrG\ (see
Table~\ref{tab:j1119_astrometry}), has $L_x/\dot E = 1\%$ (Seward
et al. 1984). \nocite{shss84} These two young but slowly spinning
pulsars therefore could have a similar efficiency for conversion
of $\dot E$ into $L_x$.

However, a very recently discovered X-ray pulsar in
SNR~G29.7$-$0.3 suggests a more complex picture. PSR~J1846$-$0258
is a rotation-powered pulsar with a period of 0.32\,s, $\tau_c =
720$\,yr, $B = 4.8 \times 10^{13}$\,G, and $\dot E = 8.3 \times
10^{36}$~\LX\ (Gotthelf et al. 2000b). \nocite{gvbt00} Given its
rotational and derived parameters, PSR~J1846$-$0258 is quite
similar to \psrG\ (Table~\ref{tab:j1119_astrometry}). Yet, for
PSR~J1846$-$0258, $L_x/\dot E = 25\%$, while $L_x/\dot E \le
0.12\%$ for \psrG. Although uncertain distances could moderate
this discrepancy, they are unlikely to solve it. The difference
could well be real, implying that the efficiency of conversion of
$\dot E$ into $L_x$ for young pulsars depends on more than just
present spin parameters --- e.g., on evolutionary state of the
pulsar--SNR system, or the local ISM conditions. It is also worth
noting that with the recent discoveries of PSRs~J1119$-$6127 and
J1846$-$0258, three of the four youngest pulsars in the Galaxy (as
ranked by characteristic age) have spin periods in the range
0.15--0.4\,s and very high inferred magnetic fields; only the Crab
spins rapidly. This emphasizes that understanding the Crab pulsar
and its nebula is not akin to understanding young pulsar--SNR
systems in general, and more sensitive follow-up studies of
systems such as \psrG--SNR~\snrG\ are essential.

\acknowledgements

Support for this work was provided by a NASA LTSA grant
(NAG5-8063), ADP grant (NAG5-9120), and NSF CAREER award
 (AST-9875897) and an NSERC Grant (RGPIN228738-00) to V.M.K.
F.~Camilo is supported by NASA grant NAG5-9095. B.M.G.
acknowledges the support of NASA through Hubble Fellowship grant
HF-01107.01-98A awarded by the Space Telescope Science Institute,
which is operated by the Association of Universities for Research
in Astronomy, Inc., for NASA under contract NAS 5-26555. This
research made use of the SIMBAD database, operated at CDS,
Strasbourg, France, as well as the HEASARC database, maintained by
NASA. We also wish to thank Mallory Roberts for his assistance
with some of the figures.

\end{document}